\documentclass[12pt]{article}

\usepackage{newtxtext,newtxmath}

\usepackage{graphicx}

\usepackage[letterpaper,margin=1in]{geometry}

\renewenvironment{abstract}
	{\quotation}
	{\endquotation}

\date{}

\makeatletter
\renewcommand{\fnum@figure}{\textbf{Figure \thefigure}}
\renewcommand{\fnum@table}{\textbf{Table \thetable}}
\makeatother

\usepackage{scicite}

\usepackage{url}

\title{Permutation invariant multi-scale full quantum neural network wavefunction}
\author{
Pengzhen~Cai$^{1\ast}$, 
Yubing~Qian$^{1,2\ast}$, 
Li~Deng$^{3,4\ast}$, 
Weizhong~Fu$^{1,2}$, 
Lei~Yang$^{4,3,5,6}$,\and
Zhiyu~Sun$^{4,3,5,6}$, 
Xin-Zheng~Li$^{1,7,8}$, 
En-Ge~Wang$^{1,7}$, 
Liangwen~Chen$^{4,3,5,6\dagger}$,\and 
Weiluo~Ren$^{2\ddagger}$,
Ji~Chen$^{1,7,8\S}$\\
\small$^{1}$School of Physics, Peking University, Beijing 100871, People's Republic of China.\and
\small$^{2}$ByteDance Seed, Beijing 100098, People's Republic of China.\and
\small$^{3}$Advanced Energy Science and Technology Guangdong Laboratory, Huizhou 516000, China.\and
\small$^{4}$Institute of Modern Physics, Chinese Academy of Sciences, Lanzhou 730000, China.\and
\small$^{5}$State Key Laboratory of Heavy Ion Science and Technology,\\
\small Institute of Modern Physics, Chinese Academy of Sciences, Lanzhou 730000, China.\and
\small$^{6}$School of Nuclear Science and Technology, University of Chinese Academy of Sciences, Beijing 100049, China.\and
\small$^{7}$Interdisciplinary Institute of Light-Element Quantum Materials,\\
\small Research Center for Light-Element Advanced Materials,\\
\small Peking University, Beijing 100871, People's Republic of China.\and
\small$^{8}$State Key Laboratory of Artificial Microstructure and Mesoscopic Physics,\\
\small Frontiers Science Center for Nano-Optoelectronics,\\
\small Peking University, Beijing 100871, People's Republic of China.\and
\small$^\ast$These authors contributed equally to this work.\and
\small$^\dagger$Corresponding author. Email: chenlw@impcas.ac.cn\and
\small$^\ddagger$Corresponding author. Email: renweiluo@bytedance.com\and
\small$^\S$Corresponding author. Email: ji.chen@pku.edu.cn.
}

\begin{document}
\maketitle

\begin{abstract}
Solving the intricate quantum behavior of interacting particles is key to unlocking the mysteries of condensed matter, but capturing their complex correlations across different scales remains a monumental challenge.
We introduce a neural network framework that overcomes this barrier by modeling the full quantum wavefunction of a system, including electrons, nuclei and muons, directly capturing the full quantum effects beyond the Born-Oppenheimer approximation. 
The neural network approximates joint wavefunction of different interacting particles with a rigorous handling of permutation invariance, enabling simultaneous treatment of nuclear quantum effects and electron-nucleus-muon couplings without explicit excited states. 
Validated on molecular systems, this approach offers a computationally feasible way to model full quantum phenomena in complex many-body systems, establishing a direct connection between fundamental particle properties and emergent material behavior.
\end{abstract}

\maketitle

\section{Introduction}
Solving the Schrödinger equation for many-particles across multiple scales holds the key to unraveling emergent phenomena in condensed matter physics and its interdisciplinary frontiers. 
However, this task remains computationally intractable due to the strong coupling between different particles, most notably electrons and nuclei, whose vastly different quantum behaviors and orders-of-magnitude mass difference present a fundamental challenge~\cite{wang_full_2026}.
The Born-Oppenheimer approximation (BOA) ~\cite{born1927born} mitigates this difficulty by decoupling nuclear and electronic motions, and has become a ubiquitous framework across diverse systems.
Within the BOA paradigm, most computational studies of molecules and materials focus either on first principles electronic structure calculations or on classical nuclear dynamics simulations. %
Nevertheless, the BOA becomes inadequate in systems involving light nucleonic particles such as light nuclei and muons, low-temperature environments, and ultrafast dynamical processes, where the intricate interplay between nucleonic and electronic degrees of freedom can no longer be neglected~\cite{wang_full_2026}.
For example, superconductivity in hydrides is strongly influenced by nuclear quantum effects~\cite{hou_effect_2025}, chemical reactions can be accelerated by nuclear quantum effects~\cite{ceriotti2016nuclear}, and the quantum fluctuation of muons can significantly influence their coupling with magnetic signals~\cite{hillier2022muon}, among others.

To capture full quantum effects beyond BOA, several methodologies have been developed, each navigating distinct trade-offs between computational scalability and accuracy.
The rigorous Born-Huang expansion (BHE)~\cite{born_dynamical_1954, zhang_ab_2018, yonehara2012fundamental} provides a formal framework for including these effects but is often computationally prohibitive due to its requirement for numerous electronic excited states.
Nuclear-electronic orbital~\cite{webb2002multiconfigurational} and multicomponent density functional theory~\cite{kreibich2001multicomponent} offer scalability, though accurate treatment of electron-nucleus correlation remains challenging.
Conversely, explicitly correlated Gaussian methods~\cite{mitroy_ecg_2013} provide benchmark accuracy yet remain restricted to small systems with few quantum nuclei.
Traditional quantum Monte Carlo approaches bridge this gap for larger systems but require specially designed trial wave functions and struggle with high electron-nuclear correlation~\cite{tubman_beyond_2014}.

\begin{figure*}
    \centering
    \includegraphics[width=\linewidth]{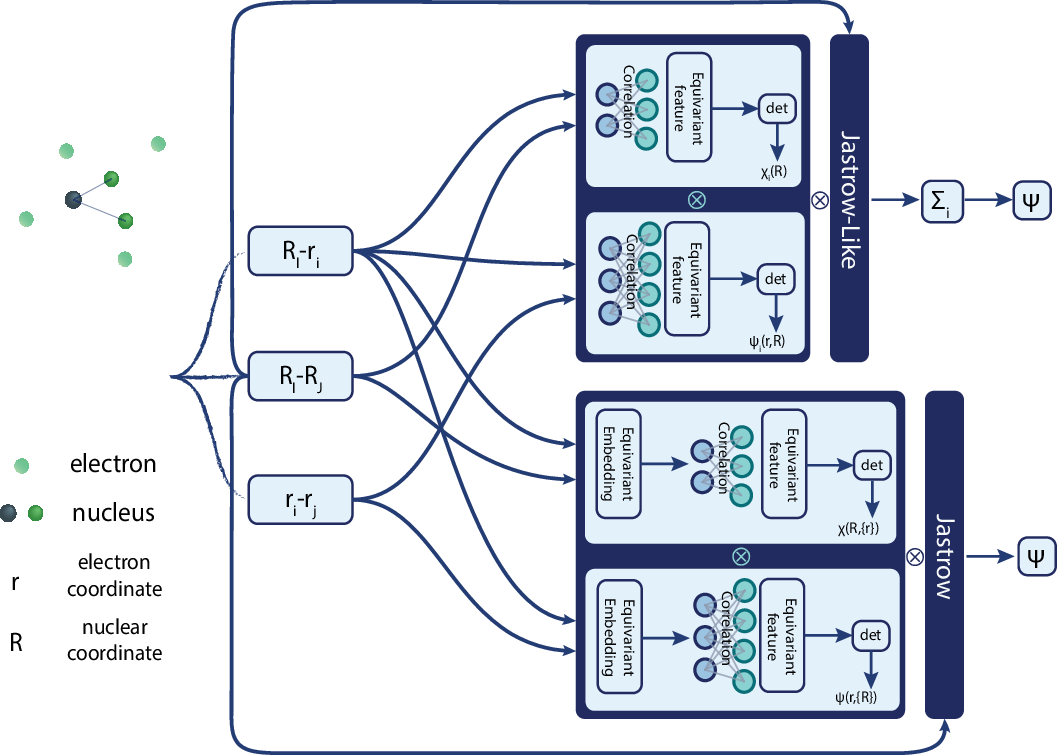}
    \caption{Sketch of neural network architecture. Due to translation invariance, only the relative positions of particles are selected as input to the neural network. Drawing on the idea of the Born-Huang expansion, the overall network can be constructed by multiplying the wavefunction components of positively and negatively charged particles (similar to how the nuclear wavefunction is multiplied by the electron wavefunction). The determinant or permanent is used to maintain exchange symmetry, and a multi-determinant/permanent structure can enhance the representation capability of neural networks.}
    \label{fig:figure1}
\end{figure*}

In this work, we introduce a permutation-invariant full-quantum neural network wavefunction architecture, dubbed PermNet, to address these challenges.
Electronic neural network wavefunction architectures have advanced rapidly over the past few years, emerging as a cutting-edge frontier in many-body electronic structure calculations~\cite{carleo_solving_2017, pfau2020ab, hermann_paulinet_2020, scherbela_deeperwin_2022, li_deepsolid_2022, ren_towards_2023, li_spin-symmetry-enforced_2024, li_lapnet_2024,  shang_solving_2025, luo_backflow_2019, hermann_nnqmc-review_2023, qian_deep_2025, tang_deep-learning_2025}.
Building on these developments, by treating electrons and nucleonic particles on an equal footing, PermNet explicitly captures electron-nucleus quantum correlations that are inherently neglected within BOA. 
Compared to other recent attempts to extend neural network wavefunctions to electron-nucleus coupled systems~\cite{zhang2025schrodingernet, chai2025revisitingbrokensymmetryphase, carr_neural_2026} and efforts to construct potential energy surfaces with a single neural network framework \cite{scherbela_deeperwin_2022, gao_pesnet_2022, gao_sampling-free_2022},
PermNet naturally enforces particle exchange symmetry, making it extendable to mixed boson-fermion systems. 
Additionally, PermNet’s unified treatment of electrons and nuclei enables the investigation of strongly correlated systems with particles of comparable masses, 
allowing for the simultaneous probing of both nuclear quantum effects and non-adiabatic effects.
Using PermNet, we demonstrate full-quantum effects of several prototypical systems, including the isotope-dependent bond lengths of hydrogen, the dielectric properties of ammonia, and the hyperfine couplings of muoniated ethylene. 
Besides, our framework facilitates the computation of the potential energy surface within BOA in a single calculation, holding promise for extension to more complex systems.

\section{Results}

\subsection{Construction of neural network wavefunction}  
The PermNet architecture (Fig.~\ref{fig:figure1}) is inspired by the BHE:  
\begin{equation}\label{eq:born-huang}  
    \Psi(\mathbf{R}, \mathbf{r})=\sum_{m}\chi_m(\mathbf{R})\psi_m(\mathbf{r};\mathbf{R}),  
\end{equation}  
where $\mathbf{r}$ and $\mathbf{R}$ denote electronic and nuclear coordinates, respectively.  
$\psi_m(\mathbf{r};\mathbf{R})$ is the $m$-th electronic state at fixed $\mathbf{R}$, and $\chi_m(\mathbf{R})$ are corresponding nuclear wave packet.  
While it is possible to compute the excited states using neural network methods~\cite{pfau_accurate_2024, li_spin-symmetry-enforced_2024, entwistle_electronic_2023}, doing so remains computationally expensive.  
Instead, we model $\psi_m^\text{net}(\mathbf{r};\mathbf{R})$ without restricting it to Hamiltonian eigenfunctions, using a general-purpose architecture such as FermiNet~\cite{pfau2020ab}:  
\begin{equation}  
    \psi_m^\text{net}(\mathbf{r};\mathbf{R})=\det[\phi^{(m)}_j(\mathbf{r}_{i},\{\mathbf{r}_{\ne i}\};\mathbf{R})].  
\end{equation}  
Here, $\mathbf{r}_i$ denotes the spatial coordinate of the $i$-th electron, and $\{\mathbf{r}_{\ne i}\}$ indicates dependence on the remaining electrons in a permutation-invariant manner, ensuring the overall antisymmetry of the electronic wavefunction.  
For $\chi_m^\text{net}(\mathbf{R})$, we generalize the Hartree product by allowing nuclear orbitals $\varphi_I(\mathbf{R})$ to depend on all nuclear coordinates:  
\begin{equation}  
    \chi_m^\text{net}(\mathbf{R})=\prod_I\varphi_I(\mathbf{R}).  
\end{equation}
The nuclear wavefunction can be further generalized to satisfy exchange symmetry while embedding the electronic coordinates, as detailed in Section~\ref{sec:nuclear-symm}.
The complete wavefunction ansatz is then given by  
\begin{equation}  
    \Psi^\text{net} = \sum_{m=1}^M\chi^\text{net}_m\psi^\text{net}_m,  
\end{equation}  
where $M$ is a tunable hyperparameter.  
Although exact factorization~\cite{abedi_exact_2010} suggests that $M=1$ is theoretically sufficient, in practice, increasing $M$ improves the accuracy of the wavefunction representation.  
Additionally, to ensure zero center-of-mass momentum in the ground state, the neural network wavefunction is constructed to be translationally invariant by using only relative coordinates, as detailed in Section~\ref{sec:internal-hamiltonian}.

\subsection{Molecular structure}

\begin{figure*}[htbp]
    \centering
    \includegraphics[width=0.9\linewidth]{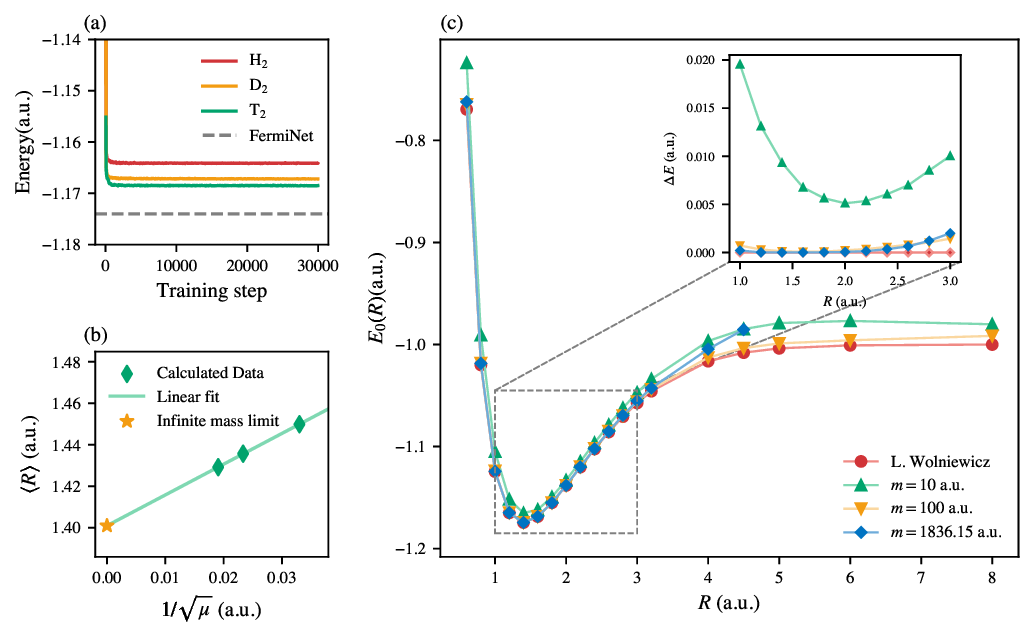}
    \caption{(a) The training curve of isotopologues of the hydrogen molecule. The Jastrow factor is chosen to be of Gaussian type. A moving window average over 500 steps is applied. The dashed line represents the energy in BOA computed by FermiNet~\cite{pfau2020ab} at equilibrium configuration. (b) The linear fit of the average internuclear distance $\langle R \rangle$ for the hydrogen molecule and its isotopologues as a function of the inverse square root of the reduced mass $1/\sqrt{\mu}$ of the two-body system. (c) PES of hydrogen calculated using a model trained under different conditions. The red line is calculated by traditional basis method~\cite{wolniewicz1995nonadiabatic}. The green, yellow and blue line are model trained under different hydrogen mass with the GEM-type Jastrow factor described in Equation~\eqref{eq:jastrow-plus}. The horizontal axis coordinate $R$ represents the static nuclear distance as in the BOA. The vertical axis of the inset $\Delta E$ represents the energy difference relative to the traditional basis method (red line). The blue line exhibits non-physical divergence at the far end and is truncated here for clarity. For details, please refer to the SI.}
    \label{fig:figure2}
\end{figure*}

The conventional notion of molecular structure as a static arrangement of nuclei at potential energy surface (PES) minima is an artificial constraint imposed by the BOA. Within our full quantum framework, molecular geometry emerges naturally from the expectation values of nuclear position operators in the entangled electron-nuclear wavefunction $\Psi(\mathbf{R},\mathbf{r})$. For any distance $R_{ij}$ between nuclei $i$ and $j$, the quantum-mechanical bond length is obtained as:
\begin{equation}
\langle R_{ij} \rangle = \frac{\int |\Psi(\mathbf{R},\mathbf{r})|^2 \, R_{ij} \, \mathrm{d}\mathbf{R}\,\mathrm{d}\mathbf{r}}{\int |\Psi(\mathbf{R},\mathbf{r})|^2 \, \mathrm{d}\mathbf{R}\,\mathrm{d}\mathbf{r}}.
\label{eq:bond_length}
\end{equation}
This formulation generalizes the classical concept of equilibrium bond length by incorporating nuclear zero-point motion and electron-nuclear correlation effects.

The distinction is most clearly illustrated in isotopologues.  
For instance, in a diatomic molecule such as the hydrogen molecule, the equilibrium distance $R_e$ predicted by BOA remains constant regardless of whether the molecule comprises hydrogen, deuterium, or tritium atoms.  
In contrast, our quantum-corrected bond length $\langle R \rangle$ deviates from the BOA result and exhibits a strong linear dependence on the inverse square root of the reduced mass $\mu$, as demonstrated in Fig.~\ref{fig:figure2}(b). The fitting yields the following relationship:
\begin{align*}
    \langle R \rangle = 1.4010 + 1.4831 \cdot \frac{1}{\sqrt{\mu}},
\end{align*}
with a coefficient of determination of 0.99993.
This result aligns with the findings of a perturbation theory analysis, detailed in the Supplemental Material, that extends beyond the lowest-order approximation of the interatomic harmonic potential by including a cubic term, thereby demonstrating that our method can effectively detect non-harmonic effects. Moreover, when extrapolated to the point corresponding to infinite mass (yellow star), this bond length shows excellent agreement with the equilibrium distance $R_e=1.400$ a.u. calculated under the BOA~\cite{martin1998benchmark}.

Under the BOA, another important concept is the potential energy surface.
Since the nuclear wave packet has a finite spatial extent, the full quantum wavefunction incorporates contributions not only from the equilibrium electronic state but also from nearby electronic states. 
If the system is well described within the BOA:  
\begin{equation}  
    \Psi(\mathbf{R},\mathbf{r}) \approx \chi_0(\mathbf{R})\psi_0(\mathbf{r};\mathbf{R}),  
\end{equation}  
then the ground state potential energy surface $E_0(\mathbf{R})$ can be extracted by integrating out the electronic degrees of freedom:  
\begin{equation}  
    E_0(\mathbf{R}) \approx \frac{\int \Psi^*(\mathbf{R},\mathbf{r}) \hat{H}_e \Psi(\mathbf{R},\mathbf{r})\, \mathrm{d}\mathbf{r}}{\int |\Psi(\mathbf{R},\mathbf{r})|^2 \, \mathrm{d}\mathbf{r}},  
\end{equation}  
where $\hat{H}_e$ denotes the electronic component of the Hamiltonian.
Unlike the traditional calculation of potential energy surfaces based on electronic wavefunctions, which necessitates optimization for each nuclear configuration and consumes substantial computational resources in high-dimensional cases, the full-quantum approach requires only a single optimization of the full quantum wavefunction.
Consequently, this method demands computational resources comparable to optimizing the electronic wavefunction at a single atomic configuration of the same system.

In Fig.~\ref{fig:figure2}(c), we present the PES calculation for the hydrogen molecule. 
A Gaussian-exponential-mixture-type (GEM-type) Jastrow factor is employed to enforce the dissociation asymptotic condition, as discussed in detail in Section \ref{sec:nuclear-jastrow}. %
The red line in Fig.~\ref{fig:figure2}(c) represents the PES of the hydrogen molecule calculated under BOA using a variational method with generalized James--Coolidge functions~\cite{wolniewicz1995nonadiabatic}. The inset of Fig.~\ref{fig:figure2}(c) shows the energy difference between the PES derived from the full quantum state and the PES calculated directly under the BOA, near the equilibrium coordinate. 
Notably, using the true nuclear mass for hydrogen, our derived PES (blue line) agrees very well with the results by Wolniewicz~\cite{wolniewicz1995nonadiabatic} (red line) near the equilibrium geometry.
However, the derived PES deviates more noticeably from the reference results at larger distances, eventually becoming too large to be displayed.
This non-physical behavior likely arises because Monte Carlo sampling is sparse in regions of small wavefunction amplitude, corresponding to nuclear coordinates far from the PES minimum.
To alleviate this error, we artificially reduce the nuclear mass to 100 a.u. (orange curve) and 10 a.u. (green curve), causing the nuclear wavefunction to become more diffused; this greatly improves sampling at long bond lengths, allowing the PES to match the reference energies more closely in that region.
However, since reducing the nuclear mass renders the BOA less valid, the PES derived from these calculations is less accurate than that obtained using the true hydrogen mass near equilibrium.
In practice, when constructing a PES using our full quantum wavefunction, one may choose an optimal effective nuclear mass to balance the numerical error induced by finite sampling against the physical bias caused by artificially small nuclear masses.

\begin{figure*}[htbp]
\centering
\includegraphics[width=\textwidth]{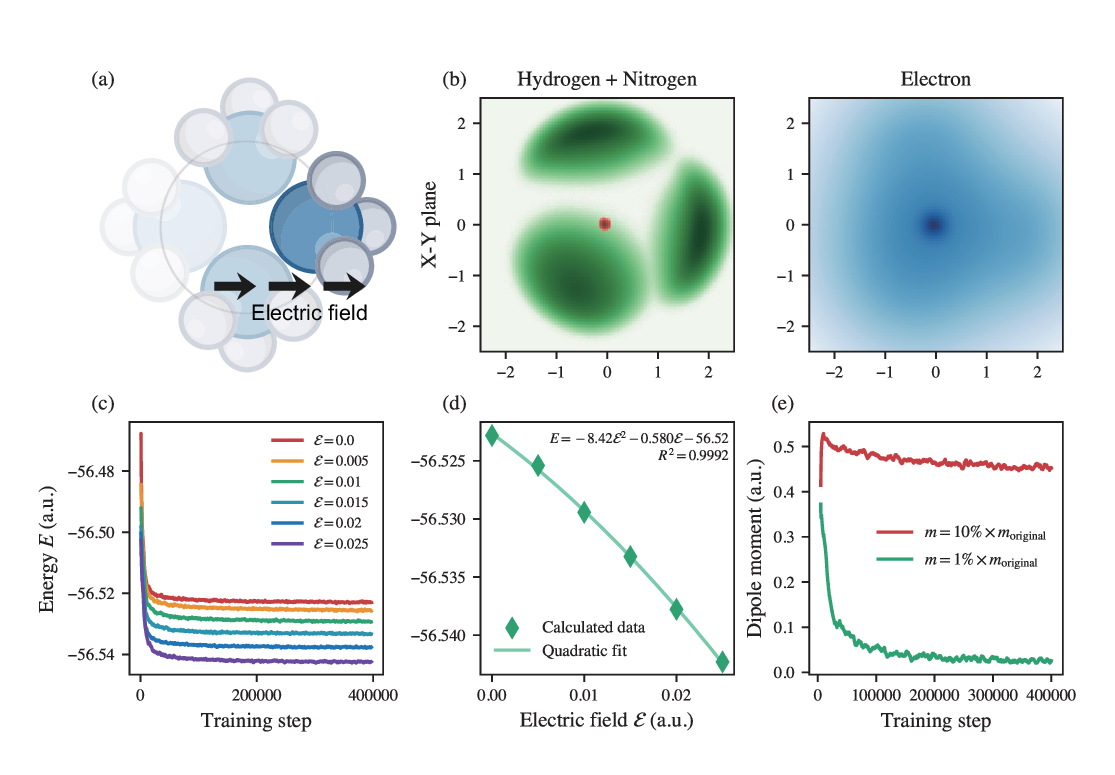}
\caption{(a) Schematic diagram of ammonia molecule polarization under an external electric field.
(b) Particle density distribution projected onto the $X$--$Y$ plane, where green, red, and blue denote hydrogen, nitrogen, and electrons, respectively, with darker shades indicating higher probability densities.
(c) Training curve of the ammonia system under uniform electric fields of varying intensities $\mathcal{E}$. A Gaussian-type Jastrow factor is used, and a moving window average over 5000 optimization steps is applied.  
(d) Quadratic fit of the ground state energy $E$ of the ammonia system as a function of the intensity $\mathcal{E}$ of the applied uniform electric field.  
(e) Dipole moment of the ammonia molecule as a function of training steps, with nuclear mass reduced to $10\%$ (red line) and $1\%$ (green line) of its original value. }
\label{fig:figure3}
\end{figure*}

\subsection{Stark effect}

Since the total Hamiltonian in full-quantum regime commutes with both the parity operator and the rotational operator, the ground state of any molecule should have zero dipole moment.
To distinguish between polar and nonpolar molecules, we rely on the Stark effect, wherein molecules with a permanent dipole moment exhibit a linear energy shift with respect to the external electric field intensity, whereas those without a permanent dipole moment display a quadratic relationship.
However, since the nuclear energy gaps are small and comparable to the energy shifts induced by the electric field, we observe a mixture of linear and quadratic Stark effects for polar molecules, as demonstrated in the following calculations using ammonia as an example. %
In Fig.~\ref{fig:figure3}(c), a series of lines illustrates the training curves for the ammonia system under an external electric field $\mathbf{\mathcal{E}}$.
These curves exhibit similar convergence rates across different electric field intensities $\mathcal{E}$ and accurately capture the ground state energy shifts induced by the field, thereby demonstrating the robustness of our method.
In Fig.~\ref{fig:figure3}(d), the energies corresponding to each electric field are computed using the well-trained wavefunction models.
We fit the relationship between the energies and electric field intensities to a quadratic curve, obtaining an excellent fit where the linear term coefficient corresponds to the permanent dipole moment $\mu=0.580\:\mathrm{a.u.}$ of the ammonia system.
This result demonstrates strong consistency with numerous experimental findings~\cite{chamberlain1969sub,marshall1981stark}.

Additionally, we directly calculated the dipole moment of the ammonia system using a model trained under zero electric field, obtaining $\mu=0.56$ a.u. in contrast to the theoretical prediction of zero.
Fig.~\ref{fig:figure3}(b) illustrates the ground-state particle distribution, offering an intuitive explanation for this non-zero dipole moment even at zero field: the hydrogen nuclei exhibit distinct localization rather than a uniform circular distribution.
This discrepancy arises because the ammonia molecule possesses an extremely small inversion energy splitting, $\Delta E = 3.6 \times 10^{-6}$\:a.u.~\cite{ho1983interstellar}, which lies beyond our current resolution limits. Due to this small energy level, minor random errors during the VMC calculation can cause symmetry breaking, leading the system to eventually converge to a localized state.
In Fig.~\ref{fig:figure3}(e), we reduce the nuclear mass to increase the energy separation. We find that when the nuclear mass is reduced to $1\%$ of its original value, the neural network can accurately distinguish the two states and provide the correct expectation value for the zero dipole moment. In addition, because ammonia has three hydrogen atoms, we test the influence of the nuclear exchange effect. As detailed in the Supplementary Material, light nuclei exhibit a significant exchange effect, which can be accounted for by enforcing nuclear exchange symmetry in PermNet. For systems with light nuclei, such as muoniated ethylene discussed in the next section, it is essential to use PermNet with nuclear exchange symmetry included.

\subsection{Muoniated molecules}

In $\mu$SR spectroscopy, positive muons are commonly employed as local magnetic probes to investigate muon-electron and muon-nuclear spin-spin interactions \cite{hillier2022muon}. For molecular systems, avoided level crossing $\mu$SR can precisely measure the hyperfine couplings of the muon, which fundamentally represent the two-body contact spin density between the muon and electrons. Given that the muon mass is only one-ninth that of a proton, properly accounting for its quantum effects is crucial. Owing to their inherent theoretical limitations, conventional methods based on density functional theory (DFT) and the BOA often fail to achieve quantitative agreement with experimental data. Within our full quantum framework, the quantum effects of muoniated molecules are captured by the expectation values of nuclear position operators in the entangled electron-nuclear wavefunction $\Psi(\mathbf{R},\mathbf{r})$.

In general, the hyperfine coupling constant $A$ for a nucleus is given by:
\begin{equation}
	A_I=\frac{2\mu_0}{3}\gamma_e\gamma_I\rho_s\left(\mathbf{r}_I\right),
\end{equation}
where $I$ denotes the nuclei and the muon, $\mu_0$ is the vacuum permeability, $\gamma$ represents the gyromagnetic ratio and $\rho_s$ is the electron-nucleus contact spin density. Regarding the hyperfine couplings of the muon, $\rho_s\left(\mathbf{r}_{\mu}\right)$ can be expressed as:

\begin{equation}
	\begin{split}
		\rho_{s}\left(\mathbf{r}_{\mu}\right) = & \left[ \sum_{i}^{N_{\uparrow}} \frac{\langle\Psi| \hat{O}_{i}^{\uparrow} \delta\left(\mathbf{r}_{i}-\mathbf{r}_{\mu}\right)|\Psi\rangle}{\langle\Psi \mid \Psi\rangle} \right. \\
		& \left. - \sum_{i}^{N_{\downarrow}} \frac{\langle\Psi| \hat{O}_{i}^{\downarrow} \delta\left(\mathbf{r}_{i}-\mathbf{r}_{\mu}\right)|\Psi\rangle}{\langle\Psi \mid \Psi\rangle} \right].
	\end{split}
\end{equation}
Here $\uparrow$ and $\downarrow$ denote spin-up and spin-down electrons, respectively, and ${\hat{O}}_i^{\uparrow/\downarrow}$ are the corresponding spin operators. Within the static muon approximation, this expression simplifies to $\rho_s(\mathbf{r}_\mu)=\rho_\uparrow(\mathbf{r}_\mu)-\rho_\downarrow(\mathbf{r}_\mu)$, where $\rho_{\uparrow/\downarrow}(\mathbf{r}_\mu)$ are the spin-up and spin-down electron densities at the muon's position.

\begin{figure*}[htbp]
\centering
\includegraphics[width=\textwidth]{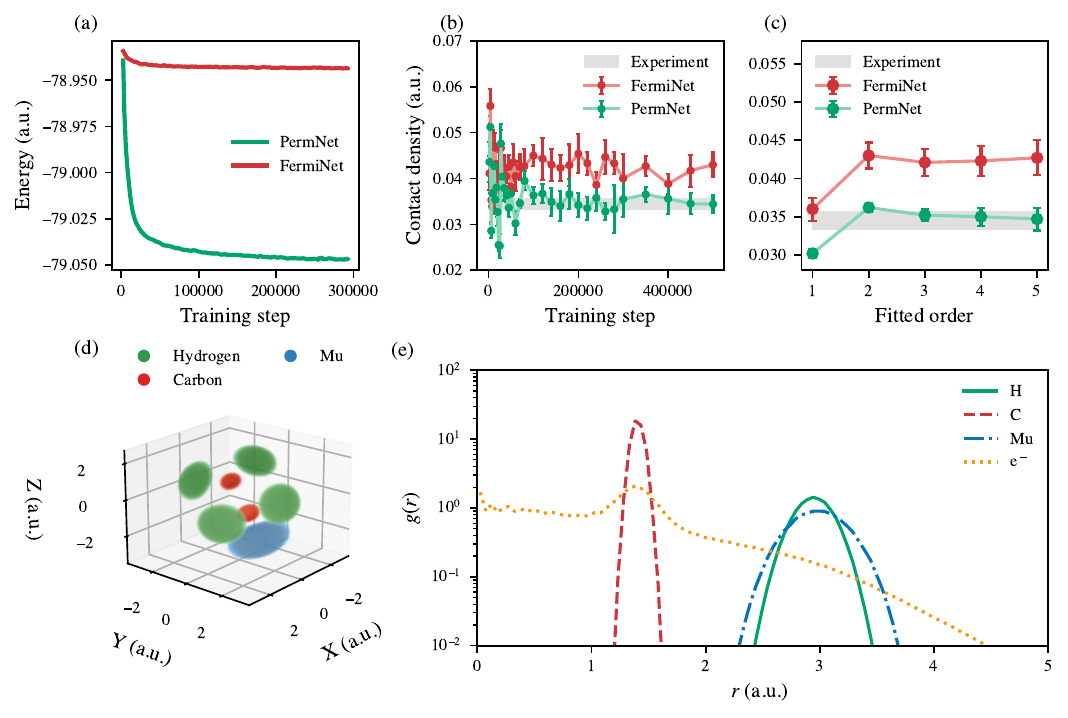}
\caption{%
(a) Training curve of the C$_2$H$_4$Mu system using PermNet (green line) and FermiNet (red line). The GEM-type Jastrow factor is used, and a moving window average over 5000 optimization steps is applied.
(b) Contact spin density as a function of training steps, calculated via the third-order polynomial extrapolation.
(c) Contact spin density calculated via different fitting order polynomial extrapolation at 500,000 training steps.
(d) Particle density distribution of the C$_2$H$_4$Mu system in real space, where green, red, and blue denote hydrogen, carbon, and muon, respectively.
(e) The radial density distributions of different particles as functions of the distance to the center of mass, where green denotes hydrogen, red dashed lines denote carbon, blue dashed lines denote the muon, and yellow dashed lines denote electrons.}
\label{fig:figure4}
\end{figure*}

In order to better understand the full quantum effects in muoniated molecules, the muoniated ethylene system (C$_2$H$_4$Mu, where Mu is the muonium) is chosen and the corresponding hyperfine coupling of the muon is investigated.
In addition to calculations performed with PermNet, where the quantum effects of all the nuclei including the muon are incorporated in the wavefunction, we also performed calculations with FermiNet, which applies the static muon approximation. The results are summarized in Fig.~\ref{fig:figure4}.

The electron-muon contact spin density is obtained by performing a histogram analysis of the muon-electron distance versus density from the wavefunction and extrapolating to zero distance (details provided in the Supplementary Material). Fig. \ref{fig:figure4}(a) shows that both PermNet and FermiNet produce converged total energies, with PermNet achieving a lower energy. This discrepancy arises because FermiNet relies on nuclear positions obtained from DFT geometry optimization, which may not represent the true energy minimum, a common challenge in calculations employing point-nucleus models.
We note that in a recent preprint that appeared during the preparation of this manuscript, Carr \textit{et al.} extended Psiformer~\cite{glehn_psiformer_2023}, a different neural network wavefunction architecture, to $\mu$SR calculations with quantum muons and protons \cite{carr_neural_2026}. 

Figs. \ref{fig:figure4}(b) and \ref{fig:figure4}(c) indicate that the calculations are converged with respect to both the number of neural network optimization steps (approximately 100,000 steps) and the order of the extrapolation polynomial (beyond the second-order). Using a third-order polynomial extrapolation and averaging results beyond 100,000 steps, PermNet yields a hyperfine coupling (represented as contact density) of 0.0348(9), while FermiNet gives 0.0429(12). The experimental reference range is 0.0334–0.0356 (150-160 MHz, extrapolation to 0 K of the muon hyperfine couplings in pure ethene) \cite{cormier2008free}. In Fig. \ref{fig:figure4}(d), the muons exhibit a wider distribution range than carbon and hydrogen, which is a natural consequence of their low mass. To enable a direct comparison of the local density, the radial density distributions of the muon, hydrogen, and carbon as functions of the distance to the center of mass are shown in Fig. \ref{fig:figure4}(e). The calculated C-Mu bond length is roughly consistent with the phonon calculation (see Supplementary Materials for details), which further supports the validity of the PermNet calculation.

PermNet demonstrates obvious superiority over FermiNet, yielding notably more reliable nuclear density distributions as well as key observables. This result establishes it as a robust method for probing full quantum effects in light nuclei, as exemplified by muoniated molecules.

\section{Discussion}

In this work, we have developed PermNet, a neural network-based quantum Monte Carlo framework that enables the direct solution of the full nuclear-electronic Schrödinger equation beyond the Born–Oppenheimer approximation. By representing the joint electron-nuclear wavefunction with a flexible neural network ansatz, our approach captures essential quantum correlations and non-adiabatic effects, offering a unified and computationally tractable route to studying coupled quantum dynamics in molecular systems. We have demonstrated the capability of this method in accurately predicting quantum-mechanical properties such as bond lengths in isotopologues, potential energy surfaces, electric field responses in molecules like ammonia, and the hyperfine couplings of muoniated molecules like ethylene, all without relying on excited electronic states calculations required by the Born–Huang formalism.

Looking forward, this approach opens promising avenues for exploring a wider range of quantum phenomena in complex systems, such as proton transfer, nuclear-quantum-effect-enhanced superconductivity, and non-adiabatic dynamics in photochemical processes, as well as the muon behavior in $\mu$SR spectroscopy. The integration of more expressive network architectures, such as symmetry-aware graph neural networks or attention mechanisms, could further enhance the representational power and physical interpretability of the wavefunction.

Despite its demonstrated accuracy, the current approach has limitations. A notable challenge is its limited resolution for rotational energy levels, as the subtle energy differences are often obscured by the statistical noise inherent in the Monte Carlo procedure. This restricts its immediate application to high-resolution rotational spectroscopy. Further developments could focus on enhancing sampling efficiency and explicitly incorporating symmetries into the ansatz to address this gap.

Nevertheless, the present work establishes a robust and extensible foundation for full quantum modeling of coupled electron-nuclear systems, bridging a critical gap between quantum theory and computational simulation in condensed matter and chemical physics.

\section{Methods}

\subsection{Translational invariance and internal Hamiltonian} \label{sec:internal-hamiltonian} 
The total Hamiltonian for a system of charged particles interacting via Coulomb forces is given by  
\begin{equation}\label{eq:full-hamiltonian}  
    \hat{H}_{\text{tot}} = -\sum_{i=1}^{N} \frac{1}{2M_i} \nabla^2_{i}  
    + \sum_{i<j}^{N} \frac{Z_i Z_j}{|\mathbf{X}_i - \mathbf{X}_j|},  
\end{equation}  
where $N$ is the total number of particles, $\mathbf{X}_i$, $M_i$, and $Z_i$ denote the spatial coordinates, mass, and charge of the $i$-th particle, and $\nabla_i^2$ is the Laplacian operator acting on the coordinates of the $i$-th particle.  
Due to the Hamiltonian's continuous translational symmetry, the ground state must have uniform density, which poses challenges in modeling the wavefunction over the entire spatial domain.  
To address this, we construct a neural network wavefunction that is explicitly translationally invariant by expressing it exclusively in terms of relative coordinates:  
\begin{equation} \label{eq12} 
    \Psi(\mathbf{X}) = \Psi(\{\mathbf{X}_i - \mathbf{X}_j\}),  
\end{equation}
where $\mathbf{X}$ represents the coordinates of all particles for brevity.
This formulation automatically removes the center-of-mass kinetic energy in Equation~\eqref{eq:full-hamiltonian}, and the remaining parts describe the internal Hamiltonian.  
This allows us to focus solely on the internal dynamics in relative coordinates, significantly simplifying the modeling of the wavefunction within a finite spatial range.  

In contrast to center-of-mass translation, rotational and vibrational motions cannot be decoupled without introducing additional approximations.

\subsection{Wavefunction optimization}

The neural network wavefunction is optimized by minimizing the variational energy of the system.  
This procedure follows the principles of traditional variational Monte Carlo, with the key distinction being the use of a neural network as the trial wavefunction.  
Given a Hamiltonian $\hat{H}$ and a neural network wavefunction $\Psi_\theta$, where $\theta$ represents the set of trainable parameters, the variational energy $E_v$ as a function of $\theta$ is expressed as
\begin{equation}
    E_v(\theta) = \frac{\int \vert\Psi_\theta(\mathbf{X})\vert^2 E_L(\mathbf{X};\theta)\,\mathrm{d}\mathbf{X}}{\int \vert\Psi_\theta(\mathbf{X})\vert^2\,\mathrm{d}\mathbf{X}}.  
\end{equation}  
The local energy $E_L(\mathbf{X};\theta)$ is defined as  
\begin{equation}  
    E_L(\mathbf{X};\theta) = \frac{\hat{H} \Psi_\theta(\mathbf{X})}{\Psi_\theta(\mathbf{X})}. 
\end{equation}  
The integrals can be efficiently evaluated using Monte Carlo integration with importance sampling, where the sampling probability is proportional to $\vert\Psi_\theta(\mathbf{X})\vert^2$.

However, the widely used Metropolis-Hastings algorithm for sampling becomes inefficient when applied to full-quantum wavefunctions.  
This inefficiency originates from two factors: different types of particles may exhibit distinct characteristic motion scales, and the full quantum wavefunction is globally translationally invariant.  
Inspired by Cassella \textit{et al.}~\cite{cassella2024neural}, we implement a modified Gibbs sampling algorithm.  
In each move proposal $\mathbf{X} \rightarrow \mathbf{X}'$, only the positions of one type of particle are updated.  
The acceptance probability for the move is given by  
\begin{equation}  
    p(\mathbf{X} \rightarrow \mathbf{X}') = \min\left\{  
        \frac{\vert\Psi_\theta(\mathbf{X}')\vert^2}{\vert\Psi_\theta(\mathbf{X})\vert^2}, 1  
    \right\}.  
\end{equation}  
This approach breaks the global translational symmetry of the trial move and allows independent adjustment of the trial move width for each particle type, thereby enhancing sampling efficiency.

We then perform gradient descent on the parameters $\theta$, using $E_v(\theta)$ as the loss function to optimize the neural network wavefunction.  
For efficient optimization, we employ the Kronecker-factored approximate curvature (KFAC)~\cite{martens_kfac_2015, dangel_kronecker-factored_2025} optimizer, which approximates the imaginary time evolution dynamics.

\subsection{Nuclear Jastrow factor}  \label{sec:nuclear-jastrow}
To improve optimization stability in the presence of limited sampling, we leverage the fact that nuclear wavefunctions are highly localized.  
We incorporate this physical insight by constraining the nuclear part of the neural network wavefunction using a Jastrow-like factor.  
As a first approximation, we assume that nucleus $I$ experiences a set of independent harmonic potentials, quadratic in the internuclear distances $R_{IJ} = |\mathbf{R}_I - \mathbf{R}_J|$.  
Accordingly, the nuclear wavefunction for nucleus $I$ is approximated as a product of Gaussian functions centered around equilibrium positions:  
\begin{equation}\label{eq:gaussian-envelope}  
    \chi(\{\mathbf{R}_I - \mathbf{R}_J\}) \propto \prod_{I,J} \exp\left[-\Gamma_{IJ} \left(R_{IJ} - R_{IJ}^{(0)}\right)^2\right],  
\end{equation}  
where $\Gamma_{IJ}$ and $R_{IJ}^{(0)}$ are trainable parameters, initialized using the internuclear distances used in the BOA.  
This Gaussian product is multiplied with the neural network nuclear wavefunction, significantly enhancing optimization stability.  

To further encode the correct asymptotic behavior when nuclei are far apart which is particularly important for PES evaluations, we extend beyond the simple Gaussian form.  
For two isolated nuclei at large separation, the wavefunction should decay exponentially with increasing internuclear distance.  
Combining this requirement with Equation~\eqref{eq:gaussian-envelope}, we construct a GEM-type Jastrow factor $\mathcal{J}$ for the nuclear orbital function as:  
\begin{equation}\label{eq:jastrow-plus}  
    \mathcal{J} = \prod_{I,J} \exp\left[-\Gamma_{IJ} \frac{\left(R_{IJ} - R_{IJ}^{(0)}\right)^2}{\sqrt{1 + \sigma_{IJ} \left(R_{IJ} - R_{IJ}^{(0)}\right)^2}}\right],  
\end{equation}  
where $\sigma_{IJ}$ is an additional learnable parameter that controls the transition between short- and long-range behavior.
The exponential decay terms between electrons and nuclei are already included in FermiNet architecture for the electrons, and we should not repeat them.

\subsection{Nuclear exchange symmetry} \label{sec:nuclear-symm}
The exchange symmetry of multi-fermion or multi-boson systems plays a crucial role in certain regimes, such as those involving low-mass particles or high-pressure conditions.  
To account for nuclear exchange symmetry, we reformulate the nuclear wavefunction by replacing the Hartree product with a determinant (for fermions) or a permanent (for bosons):  
\begin{equation}\label{eq:symmetric-nuc-wf}  
    \chi_m^\text{net} = \prod_\alpha \det\left[\varphi_{K}^{(m)}(\mathbf{R}_{\alpha,L},\{\mathbf{R}_{\alpha,\ne L}\},\{\mathbf{R}_{\ne\alpha}\};\mathbf{r})\right],
\end{equation}
where $K$ labels orbitals and $L$ labels nuclei within species $\alpha$, and they correspond to the rows and columns of the determinant, respectively.  
For bosonic species, the determinant is replaced by a permanent to ensure the correct exchange symmetry under particle permutation.

Most previous neural network architectures address permutation symmetry for only one type of particle.  
These approaches are insufficient in the full-quantum case, where we must ensure that permuting electrons does not affect the nuclear part of the wavefunction, and vice versa.  
Inspired by Cassella \textit{et al.} in the context of positronic chemistry, we propose a Deep Sets-like~\cite{zaheer2017deep} structure for constructing input features.  
The single-particle input features are divided into positive ($\mathbf{h}^{(+)}$) and negative ($\mathbf{h}^{(-)}$) charge channels:  
\begin{align}  
    h_{Ip}^{(+)} = \sum_{i} \mathbf{W}_{Ip} \mathcal{D}(\mathbf{R}_I, \mathbf{r}_i) + b_{Ip}, \\  
    h_{ip}^{(-)} = \sum_{I} \mathbf{V}_{ip} \mathcal{D}(\mathbf{r}_i, \mathbf{R}_I) + c_{ip},
\end{align}  
where $h_{Ip}^{(+)}$ is the $p$-th feature of nucleus $I$'s one-particle feature, and similar for $h_{ip}^{(-)}$.
$\mathbf{W}_{Ip}$, $\mathbf{V}_{ip}$, $b_{Ip}$, and $c_{ip}$ are learnable parameters.
$\mathcal{D}$ is a pairwise descriptor function that takes two position vectors and returns the concatenated scalar distance and displacement vector:  
\begin{equation}  
    \mathcal{D}(\mathbf{X}_i, \mathbf{X}_j) = \begin{pmatrix}  
        \vert \mathbf{X}_i - \mathbf{X}_j \vert \\  
        \mathbf{X}_i - \mathbf{X}_j  
    \end{pmatrix} \in \mathbb{R}^4.  
\end{equation}  
Similarly, the two-particle input features are constructed as:  
\begin{align}  
    \mathbf{g}_{IJ}^{(+)} &= \text{concatenate}\left[\mathcal{D}(\mathbf{r}_I, \mathbf{r}_J); \forall J\right], \\  
    \mathbf{g}_{ij}^{(-)} &= \text{concatenate}\left[\mathcal{D}(\mathbf{r}_i, \mathbf{r}_j); \forall j\right].  
\end{align}

\bibliographystyle{sciencemag}
\bibliography{references}

\begin{thebibliography}{0}%
\makeatletter
\providecommand \@ifxundefined [1]{%
 \@ifx{#1\undefined}
}%
\providecommand \@ifnum [1]{%
 \ifnum #1\expandafter \@firstoftwo
 \else \expandafter \@secondoftwo
 \fi
}%
\providecommand \@ifx [1]{%
 \ifx #1\expandafter \@firstoftwo
 \else \expandafter \@secondoftwo
 \fi
}%
\providecommand \natexlab [1]{#1}%
\providecommand \enquote  [1]{``#1''}%
\providecommand \bibnamefont  [1]{#1}%
\providecommand \bibfnamefont [1]{#1}%
\providecommand \citenamefont [1]{#1}%
\providecommand \href@noop [0]{\@secondoftwo}%
\providecommand \href [0]{\begingroup \@sanitize@url \@href}%
\providecommand \@href[1]{\@@startlink{#1}\@@href}%
\providecommand \@@href[1]{\endgroup#1\@@endlink}%
\providecommand \@sanitize@url [0]{\catcode `\\12\catcode `\$12\catcode
  `\&12\catcode `\#12\catcode `\^12\catcode `\_12\catcode `\%12\relax}%
\providecommand \@@startlink[1]{}%
\providecommand \@@endlink[0]{}%
\providecommand \url  [0]{\begingroup\@sanitize@url \@url }%
\providecommand \@url [1]{\endgroup\@href {#1}{\urlprefix }}%
\providecommand \urlprefix  [0]{URL }%
\providecommand \Eprint [0]{\href }%
\providecommand \doibase [0]{https://doi.org/}%
\providecommand \selectlanguage [0]{\@gobble}%
\providecommand \bibinfo  [0]{\@secondoftwo}%
\providecommand \bibfield  [0]{\@secondoftwo}%
\providecommand \translation [1]{[#1]}%
\providecommand \BibitemOpen [0]{}%
\providecommand \bibitemStop [0]{}%
\providecommand \bibitemNoStop [0]{.\EOS\space}%
\providecommand \EOS [0]{\spacefactor3000\relax}%
\providecommand \BibitemShut  [1]{\csname bibitem#1\endcsname}%
\let\auto@bib@innerbib\@empty
\end{thebibliography}%


\begin{thebibliography}{10}
\providecommand{\url}[1]{\texttt{#1}}
\expandafter\ifx\csname urlstyle\endcsname\relax
  \providecommand{\doi}[1]{doi:\discretionary{}{}{}#1}\else
  \providecommand{\doi}{doi:\discretionary{}{}{}\begingroup \urlstyle{rm}\Url}\fi

\bibitem{wang_full_2026}
E.~Wang, \emph{Full {{Quantum Effects}} in {{Condensed Matter Physics}}: {{Beyond}} the {{Born-Oppenheimer Approximation}}} (Cambridge University Press, Cambridge) (2026).

\bibitem{born1927born}
M.~Born, R.~Oppenheimer, Zur Quantentheorie Der Molekeln. \emph{Annalen der Physik} \textbf{389}~(20), 457--484 (1927).

\bibitem{hou_effect_2025}
P.~Hou, F.~Belli, T.~Bi, E.~Zurek, I.~Errea, Effect of Quantum Anharmonicity on the Superconductivity of {$I\overline{4}3m$} {${\mathrm{CH}}_{4}\text{\ensuremath{-}}{\mathrm{H}}_{3}\mathrm{S}$} at High Pressures: {{A}} First-Principles Study. \emph{Physical Review B} \textbf{111}~(14), 144509 (2025).

\bibitem{ceriotti2016nuclear}
M.~Ceriotti, \emph{et~al.}, Nuclear quantum effects in water and aqueous systems: Experiment, theory, and current challenges. \emph{Chemical reviews} \textbf{116}~(13), 7529--7550 (2016).

\bibitem{hillier2022muon}
A.~D. Hillier, \emph{et~al.}, Muon spin spectroscopy. \emph{Nature Reviews Methods Primers} \textbf{2}~(1), 4 (2022).

\bibitem{born_dynamical_1954}
M.~Born, K.~Huang, \emph{Dynamical Theory of Crystal Lattices} (Clarendon Press, Oxford) (1954).

\bibitem{zhang_ab_2018}
X.-W. Zhang, E.-G. Wang, X.-Z. Li, Ab Initio Investigation on the Experimental Observation of Metallic Hydrogen. \emph{Physical Review B} \textbf{98}~(13), 134110 (2018), \doi{10.1103/PhysRevB.98.134110}.

\bibitem{yonehara2012fundamental}
T.~Yonehara, K.~Hanasaki, K.~Takatsuka, Fundamental approaches to nonadiabaticity: Toward a chemical theory beyond the {B}orn--{O}ppenheimer paradigm. \emph{Chemical Reviews} \textbf{112}~(1), 499--542 (2012).

\bibitem{webb2002multiconfigurational}
S.~P. Webb, T.~Iordanov, S.~Hammes-Schiffer, Multiconfigurational nuclear-electronic orbital approach: Incorporation of nuclear quantum effects in electronic structure calculations. \emph{The Journal of chemical physics} \textbf{117}~(9), 4106--4118 (2002).

\bibitem{kreibich2001multicomponent}
T.~Kreibich, E.~Gross, Multicomponent density-functional theory for electrons and nuclei. \emph{Physical Review Letters} \textbf{86}~(14), 2984 (2001).

\bibitem{mitroy_ecg_2013}
J.~Mitroy, \emph{et~al.}, Theory and Application of Explicitly Correlated {{Gaussians}}. \emph{Reviews of Modern Physics} \textbf{85}~(2), 693--749 (2013), \doi{10.1103/RevModPhys.85.693}.

\bibitem{tubman_beyond_2014}
N.~M. Tubman, I.~Kyl{\"a}np{\"a}{\"a}, S.~{Hammes-Schiffer}, D.~M. Ceperley, Beyond the {{Born-Oppenheimer}} Approximation with Quantum {{Monte Carlo}} Methods. \emph{Physical Review A} \textbf{90}~(4), 042507 (2014), \doi{10.1103/PhysRevA.90.042507}.

\bibitem{carleo_solving_2017}
G.~Carleo, M.~Troyer, Solving the Quantum Many-Body Problem with Artificial Neural Networks. \emph{Science} \textbf{355}~(6325), 602--606 (2017).

\bibitem{pfau2020ab}
D.~Pfau, J.~S. Spencer, A.~G. Matthews, W.~M.~C. Foulkes, Ab initio solution of the many-electron Schr{\"o}dinger equation with deep neural networks. \emph{Physical review research} \textbf{2}~(3), 033429 (2020).

\bibitem{hermann_paulinet_2020}
J.~Hermann, Z.~Sch{\"a}tzle, F.~No{\'e}, Deep-Neural-Network Solution of the Electronic {{Schr\"odinger}} Equation. \emph{Nature Chemistry} \textbf{12}~(10), 891--897 (2020).

\bibitem{scherbela_deeperwin_2022}
M.~Scherbela, R.~Reisenhofer, L.~Gerard, P.~Marquetand, P.~Grohs, Solving the Electronic {{Schr\"odinger}} Equation for Multiple Nuclear Geometries with Weight-Sharing Deep Neural Networks. \emph{Nature Computational Science} \textbf{2}~(5), 331--341 (2022).

\bibitem{li_deepsolid_2022}
X.~Li, Z.~Li, J.~Chen, Ab Initio Calculation of Real Solids via Neural Network Ansatz. \emph{Nature Communications} \textbf{13}~(1), 7895 (2022).

\bibitem{ren_towards_2023}
W.~Ren, W.~Fu, X.~Wu, J.~Chen, Towards the Ground State of Molecules via Diffusion {{Monte Carlo}} on Neural Networks. \emph{Nature Communications} \textbf{14}~(1), 1860 (2023).

\bibitem{li_spin-symmetry-enforced_2024}
Z.~Li, \emph{et~al.}, Spin-Symmetry-Enforced Solution of the Many-Body {{Schr\"odinger}} Equation with a Deep Neural Network. \emph{Nature Computational Science} \textbf{4}~(12), 910--919 (2024).

\bibitem{li_lapnet_2024}
R.~Li, \emph{et~al.}, A Computational Framework for Neural Network-Based Variational {{Monte Carlo}} with {{Forward Laplacian}}. \emph{Nature Machine Intelligence} \textbf{6}~(2), 209--219 (2024), \doi{10.1038/s42256-024-00794-x}.

\bibitem{shang_solving_2025}
H.~Shang, C.~Guo, Y.~Wu, Z.~Li, J.~Yang, Solving the Many-Electron {{Schr\"odinger}} Equation with a Transformer-Based Framework. \emph{Nature Communications} \textbf{16}~(1), 8464 (2025).

\bibitem{luo_backflow_2019}
D.~Luo, B.~K. Clark, Backflow {{Transformations}} via {{Neural Networks}} for {{Quantum Many-Body Wave Functions}}. \emph{Physical Review Letters} \textbf{122}~(22), 226401 (2019).

\bibitem{hermann_nnqmc-review_2023}
J.~Hermann, \emph{et~al.}, Ab Initio Quantum Chemistry with Neural-Network Wavefunctions. \emph{Nature Reviews Chemistry} \textbf{7}~(10), 692--709 (2023).

\bibitem{qian_deep_2025}
Y.~Qian, X.~Li, Z.~Li, W.~Ren, J.~Chen, Deep {{Learning Quantum Monte Carlo}} for {{Solids}}. \emph{WIREs Computational Molecular Science} \textbf{15}~(2), e70015 (2025).

\bibitem{tang_deep-learning_2025}
Z.~Tang, \emph{et~al.}, Deep-Learning Electronic Structure Calculations. \emph{Nature Computational Science} \textbf{5}~(12), 1133--1146 (2025).

\bibitem{zhang2025schrodingernet}
Y.~Zhang, B.~Jiang, H.~Guo, SchrodingerNet: A Universal Neural Network Solver for the Schrodinger Equation. \emph{Journal of Chemical Theory and Computation} \textbf{21}~(2), 670--677 (2025).

\bibitem{chai2025revisitingbrokensymmetryphase}
S.~Chai, \emph{et~al.}, Revisiting the Broken Symmetry Phase of Solid Hydrogen: A Neural Network Variational Monte Carlo Study (2025).

\bibitem{carr_neural_2026}
J.~Carr, M.~Volkai, W.~M.~C. Foulkes, A.~P. Fadon, Neural {{Wavefunction Calculations}} of {{$\mu$SR Spectra}} with {{Quantum Muons}} and {{Protons}} (2026).

\bibitem{gao_pesnet_2022}
N.~Gao, S.~G{\"u}nnemann, Ab-{{Initio Potential Energy Surfaces}} by {{Pairing GNNs}} with {{Neural Wave Functions}}, in \emph{International {{Conference}} on {{Learning Representations}}} (arXiv) (2021).

\bibitem{gao_sampling-free_2022}
N.~Gao, S.~G{\"u}nnemann, Sampling-Free {{Inference}} for {{Ab-Initio Potential Energy Surface Networks}}, in \emph{The {{Eleventh International Conference}} on {{Learning Representations}}} (2022).

\bibitem{pfau_accurate_2024}
D.~Pfau, S.~Axelrod, H.~Sutterud, I.~Von~Glehn, J.~S. Spencer, Accurate Computation of Quantum Excited States with Neural Networks. \emph{Science} \textbf{385}~(6711), eadn0137 (2024).

\bibitem{entwistle_electronic_2023}
M.~T. Entwistle, Z.~Sch{\"a}tzle, P.~A. Erdman, J.~Hermann, F.~No{\'e}, Electronic Excited States in Deep Variational {{Monte Carlo}}. \emph{Nature Communications} \textbf{14}~(1), 274 (2023).

\bibitem{abedi_exact_2010}
A.~Abedi, N.~T. Maitra, E.~K.~U. Gross, Exact {{Factorization}} of the {{Time-Dependent Electron-Nuclear Wave Function}}. \emph{Physical Review Letters} \textbf{105}~(12), 123002 (2010).

\bibitem{wolniewicz1995nonadiabatic}
L.~Wolniewicz, Nonadiabatic energies of the ground state of the hydrogen molecule. \emph{The Journal of chemical physics} \textbf{103}~(5), 1792--1799 (1995).

\bibitem{martin1998benchmark}
J.~M. Martin, Benchmark ab initio potential curves for the light diatomic hydrides. Unusually large nonadiabatic effects in BeH and BH. \emph{Chemical physics letters} \textbf{283}~(5-6), 283--293 (1998).

\bibitem{chamberlain1969sub}
J.~Chamberlain, A.~Costley, H.~Gebbie, The sub-millimetre dispersion, rotational line strengths and dipole moment of gaseous ammonia. \emph{Spectrochimica Acta Part A: Molecular Spectroscopy} \textbf{25}~(1), 9--18 (1969).

\bibitem{marshall1981stark}
M.~D. Marshall, J.~S. Muenter, Stark-Hyperfine Measurements in {{{\emph{K}}}} = 0 and {{{\emph{K}}}} = 1 States of {{NH3}}. \emph{Journal of Molecular Spectroscopy} \textbf{85}~(2), 322--326 (1981).

\bibitem{ho1983interstellar}
P.~T. Ho, C.~H. Townes, Interstellar ammonia. \emph{IN: Annual review of astronomy and astrophysics. Volume 21 (A84-10851 01-90). Palo Alto, CA, Annual Reviews, Inc., 1983, p. 239-270.} \textbf{21}, 239--270 (1983).

\bibitem{glehn_psiformer_2023}
I.~{von Glehn}, J.~S. Spencer, D.~Pfau, A Self-Attention Ansatz for Ab-Initio Quantum Chemistry, in \emph{The Eleventh International Conference on Learning Representations, {{ICLR}} 2023} (OpenReview.net, kigali, rwanda) (2023).

\bibitem{cormier2008free}
P.~Cormier, \emph{et~al.}, Free radical formation in supercritical CO2, using muonium as a probe and implication for H atom reaction with ethene. \emph{The Journal of Physical Chemistry A} \textbf{112}~(20), 4593--4600 (2008).

\bibitem{cassella2024neural}
G.~Cassella, W.~Foulkes, D.~Pfau, J.~S. Spencer, Neural network variational {M}onte {C}arlo for positronic chemistry. \emph{Nature Communications} \textbf{15}~(1), 5214 (2024).

\bibitem{martens_kfac_2015}
J.~Martens, R.~Grosse, Optimizing Neural Networks with {K}ronecker-factored Approximate Curvature, in \emph{Proceedings of the 32nd International Conference on Machine Learning} (PMLR) (2015), pp. 2408--2417.

\bibitem{dangel_kronecker-factored_2025}
F.~Dangel, B.~Mucs{\'a}nyi, T.~Weber, R.~Eschenhagen, Kronecker-Factored {{Approximate Curvature}} ({{KFAC}}) {{From Scratch}} (2025).

\bibitem{zaheer2017deep}
M.~Zaheer, \emph{et~al.}, Deep sets. \emph{Advances in neural information processing systems} \textbf{30} (2017).

\end{thebibliography}

\section*{Acknowledgement}
The authors thank the High Performance Computing Platform of Peking University for computational resources and the Dongjiang Yuan Intelligent Computing Center supercomputing facility.

\paragraph*{Funding:}
J.C. is funded by the National Science Foundation of China under Grant No. 12334003.
Y.Q. is supported by the National Natural Science Foundation of China under Grant No. 125B2083.
L.C. is supported by the National Natural Science Foundation of China under Grant No. 12405337,
the Guangdong Basic and Applied Basic Research Foundation under Grant No. 2023B1515120067,
the Research Program of State Key Laboratory of Heavy Ion Science and Technology, Institute of Modern Physics, Chinese Academy of Sciences, under Grant No. HIST2025CS06.
X.L. is supported by the National Natural Science Foundation of China under Grant No. 12550005.
\paragraph*{Author contributions:}
Conceptualization: L.C., W.R., and J.C.
Data curation: P.C., Y.Q., and L.D.
Formal analysis: P.C., Y.Q., and L.D.
Funding acquisition: Y.Q., X.L., L.C., W.R., and J.C.
Investigation: P.C., Y.Q., and L.D.
Methodology: P.C., Y.Q., L.D., and W.F.
Project administration: L.C., W.R., and J.C.
Resources: L.Y., Z.S., X.L., E.W., L.C., W.R., and J.C.
Software: P.C., Y.Q., and L.D.
Supervision: L.Y., Z.S., X.L., E.W., L.C., W.R., and J.C.
Validation: P.C., Y.Q., L.D., L.C., W.R., and J.C.
Visualization: P.C., Y.Q., and L.D.
Writing—original draft:  P.C., Y.Q., L.D., L.C., W.R., and J.C.
Writing—review and editing: P.C., Y.Q., L.D., W.F., L.Y., Z.S., X.L., E.W., L.C., W.R., and J.C.
\paragraph*{Competing interests:}
There are no competing interests to declare.
\paragraph*{Data and materials availability:}
All data underlying the study, including the neural network parameters and training scripts, will be deposited in a public repository (e.g., Zenodo) upon publication of the manuscript. Currently, the datasets are available from the corresponding authors upon reasonable request. There are no restrictions on data availability, and no Materials Transfer Agreements (MTA) are required.

\end{document}


\title{Supplementary material: Permutation invariant multi-scale full quantum neural network wavefunction}

\author{Pengzhen Cai}
\thanks{These authors contributed equally to this work.}
\affiliation{School of Physics, Peking University, Beijing 100871, People's Republic of China}

\author{Yubing Qian}
\thanks{These authors contributed equally to this work.}
\affiliation{School of Physics, Peking University, Beijing 100871, People's Republic of China}
\affiliation{ByteDance Seed, Beijing 100098, People’s Republic of China}

\author{Li Deng}
\thanks{These authors contributed equally to this work.}
\affiliation{Advanced Energy Science and Technology Guangdong Laboratory, Huizhou 516000, China}	
\affiliation{Institute of Modern Physics, Chinese Academy of Sciences, Lanzhou 730000, China}

\author{Weizhong Fu}
\affiliation{School of Physics, Peking University, Beijing 100871, People's Republic of China}
\affiliation{ByteDance Seed, Beijing 100098, People’s Republic of China}

\author{Lei Yang}
\affiliation{Institute of Modern Physics, Chinese Academy of Sciences, Lanzhou 730000, China}
\affiliation{Advanced Energy Science and Technology Guangdong Laboratory, Huizhou 516000, China}	
\affiliation{State Key Laboratory of Heavy Ion Science and Technology, Institute of Modern Physics, Chinese
Academy of Sciences, Lanzhou 730000, China}
\affiliation{School of Nuclear Science and Technology, University of Chinese Academy of Sciences, Beijing 100049, China}

\author{Zhiyu Sun}
\affiliation{Institute of Modern Physics, Chinese Academy of Sciences, Lanzhou 730000, China}
\affiliation{Advanced Energy Science and Technology Guangdong Laboratory, Huizhou 516000, China}	
\affiliation{State Key Laboratory of Heavy Ion Science and Technology, Institute of Modern Physics, Chinese
Academy of Sciences, Lanzhou 730000, China}
\affiliation{School of Nuclear Science and Technology, University of Chinese Academy of Sciences, Beijing 100049, China}

\author{Xin-Zheng Li}
\affiliation{School of Physics, Peking University, Beijing 100871, People's Republic of China}
\affiliation{Interdisciplinary Institute of Light-Element Quantum Materials, Research Center for Light-Element Advanced Materials, Peking University, Beijing 100871, People's Republic of China}
\affiliation{State Key Laboratory of Artificial Microstructure and Mesoscopic Physics, Frontiers Science Center for Nano-Optoelectronics, Peking University, Beijing 100871, People's Republic of China}

\author{En-Ge Wang}
\affiliation{School of Physics, Peking University, Beijing 100871, People's Republic of China}
\affiliation{Interdisciplinary Institute of Light-Element Quantum Materials, Research Center for Light-Element Advanced Materials, Peking University, Beijing 100871, People's Republic of China}

\author{Liangwen Chen}
\email{chenlw@impcas.ac.cn}
\affiliation{Institute of Modern Physics, Chinese Academy of Sciences, Lanzhou 730000, China}
\affiliation{Advanced Energy Science and Technology Guangdong Laboratory, Huizhou 516000, China}	
\affiliation{State Key Laboratory of Heavy Ion Science and Technology, Institute of Modern Physics, Chinese
Academy of Sciences, Lanzhou 730000, China}
\affiliation{School of Nuclear Science and Technology, University of Chinese Academy of Sciences, Beijing 100049, China}

\author{Weiluo Ren}
\email{renweiluo@bytedance.com}
\affiliation{ByteDance Seed, Beijing 100098, People’s Republic of China}

\author{Ji Chen}
\email{ji.chen@pku.edu.cn}
\affiliation{School of Physics, Peking University, Beijing 100871, People's Republic of China}
\affiliation{Interdisciplinary Institute of Light-Element Quantum Materials, Research Center for Light-Element Advanced Materials, Peking University, Beijing 100871, People's Republic of China}
\affiliation{State Key Laboratory of Artificial Microstructure and Mesoscopic Physics, Frontiers Science Center for Nano-Optoelectronics, Peking University, Beijing 100871, People's Republic of China}

\maketitle

\tableofcontents

\clearpage
\section{Asymptotic Jastrow factors}
\begin{figure}[htbp]
\centering
\includegraphics[width=0.6\linewidth]{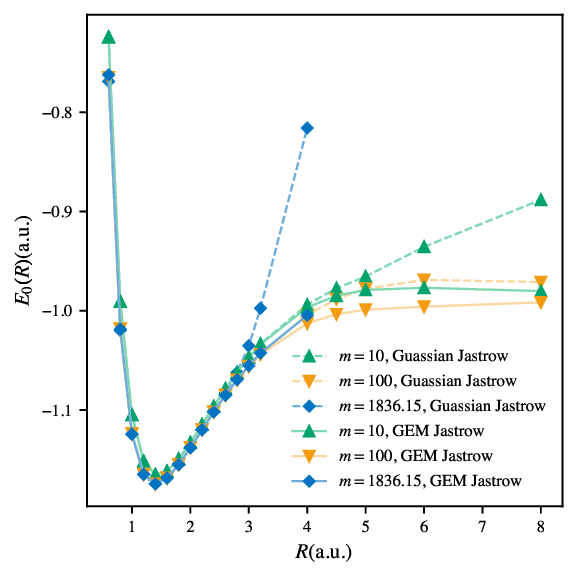}
\caption{PES of hydrogen calculated using a model trained under different conditions. The green, yellow and blue line are model trained under different hydrogen mass. The dashed lines represent the model with a Gaussian-type Jastrow factor, while the solid lines represent the model with the GEM-type Jastrow factor described in the main text.}
\label{fig:s1}
\end{figure}

The increasing errors in regions of PES away from the equilibrium point when using a full quantum state to calculate is inherent because of the important sampling. As designed in the main text, we exhibit the effect of asymptotic Jastrow factors on PES calculation comparing to the simple Gaussian-type Jastrow factors in Fig.~\ref{fig:s1}

\clearpage
\section{Nuclear exchange effect}
The key challenge in enforcing nuclear exchange symmetry is the inclusion of different particles in a single function, such as an electronic wavefunction that contains both electronic and nuclear coordinates, $\psi_e(\mathbf{r},\mathbf{R})$. Since nuclei can be fermions or bosons, a straightforward strategy is to isolate and handle the exchange symmetry within each function, such that:
\begin{equation}
    \psi_e(\mathbf{r},\mathbf{R})\rightarrow\psi_e(\mathbf{r},\{\mathbf{R}\})
\end{equation}
The detailed implementation of this property is explained in the main text. The exchange symmetry can then be treated separately in the nuclear and electronic wave functions. Alternatively, one can embed the electronic coordinates into the nuclear wave function:
\begin{equation}\label{eq:wf-embeding}
    \chi(\mathbf{R})\rightarrow\chi(\mathbf{R},\{\mathbf{r}\})
\end{equation}

\begin{figure}[htbp]
\centering
\includegraphics[width=0.9\linewidth]{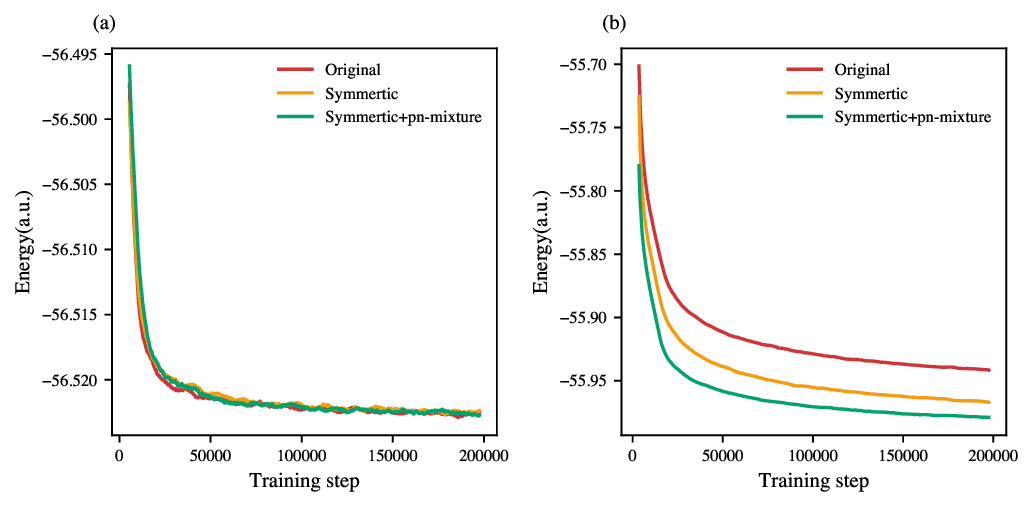}
\caption{(a) Training curve of the ammonia system. The red line represents the Hatree-type nuclear wavefunction, which has no constraints on nuclear exchange symmetry. The orange line represents the nuclear exchange symmetry enforcing wavefunction, but the nuclear wavefunction contains only nuclear coordinates. The green line represents the nuclear exchange symmetry enforcing wavefunction, with the nuclear wavefunction containing electronic coordinates in addition, as expressed in Equation~\eqref{eq:wf-embeding}. (b) Training curve of the ammonia system with nuclear mass reduced to $1\%$ of its original value. All other settings match those in panel (a).}
\label{fig:s2}
\end{figure}

The results in Fig.~\ref{fig:s2} indicate that for systems with light nuclei, it is essential to account for nuclear exchange symmetry. In such cases, the mixing of nuclear and electronic coordinates in the nuclear wavefunction can further enhance the expressive power of PermNet.
\clearpage
\section{Analysis of the relationship between bond length and nuclear mass using perturbation theory}
In this section, we present a simple calculation of the relationship between bond length and nuclear mass in a diatomic molecule, using atomic units. This relationship can be derived by considering the lowest-order anharmonic term in the potential energy. Within the BOA, the nuclear wavefunction of a diatomic molecule satisify:
\begin{equation}
    \left(-\frac{\nabla_1^2}{2M_1}-\frac{\nabla_2^2}{2M_2}+E_0(\mathbf{R}_1,\mathbf{R}_2)\right)\chi(\mathbf{R}_1,\mathbf{R}_2)=E\chi(\mathbf{R}_1,\mathbf{R}_2)
\end{equation}
where $m_1$ and $m_2$ denote the nuclear masses of the two nuclei in a diatomic molecule, $\mathbf{R}_1$ and $\mathbf{R}_2$ denote their coordinates, and $\nabla_1^2$ and $\nabla_2^2$ are the Laplacian operators acting on $\mathbf{R}_1$ and $\mathbf{R}_2$, respectively. Here, $E_0$ is the PES and $E$ is a constant. Considering the two-body reduction, we introduce $\mathbf{R}=\mathbf{R}_1-\mathbf{R}_2$ as the only variable, with $\mu=\sqrt{\frac{M_1M_2}{M_1+M_2}}$ as the reduced mass, and neglect the irrelevant center-of-mass and rotational motions. The equation can then be transformed into:
\begin{equation}
    \left(-\frac{1}{2\mu}\frac{\partial^2}{\partial R^2}+E_0(R)\right)\chi(R)=E\chi(R)
\end{equation}
Around the equilibrium distance $R_0$, the potential energy can be expanded as:
\begin{equation}
    E_0(R)=E_0^0+a_2(R-R_0)^2+a_3(R-R_0)^3+\cdots
\end{equation}
where $E_0^0$, $a_2$, and $a_3$ are constants. The energy shift $E_0^0$ can be absorbed into $E$, and the harmonic term dominates the potential. Treating the cubic term as a perturbation, the ground-state nuclear wavefunction near harmonic states can be approximated as $\ket{\psi}=\ket{\psi^{(0)}}+\ket{\psi^{(1)}}+...$. Since the states can be chosen as real, the bond length is then calculated as:
\begin{equation}\label{eq:bond_length}
\langle R\rangle=\langle \psi| R|\psi\rangle=\langle \psi^{(0)}| R|\psi^{(0)}\rangle + 2\langle \psi^{(1)}| R|\psi^{(0)}\rangle
\end{equation}
The ground harmonic state $\ket{\psi_0}=\ket{0}$ is symmetric about the equilibrium point, so $\langle \psi_0| R|\psi_0\rangle=R_0$. Define the harmonic frequency $\omega=\sqrt{\frac{2a_2}{\mu}}$. According to perturbation theory, the first-order perturbation term is:
\begin{equation}\label{eq:perturbation_term}
    \ket{\psi^{(1)}}=\sum_{m\neq 0}\frac{\langle m|a_3(R-R_0)^3|0\rangle}{(0-m)\omega}\ket{m}
\end{equation}
where $\ket{m}$ denotes the $m$-th excited harmonic state. Using the orthogonality between $\ket{\psi^{(0)}}$ and $\ket{\psi^{(1)}}$, we have $\langle\psi^{(1)}|R_0|\psi^{(0)}\rangle=0$. Substituting Equation \ref{eq:perturbation_term} into Equation \ref{eq:bond_length} yields:
\begin{equation}
    \langle R\rangle=R_0-\sum_{m\neq 0}\frac{\langle 0|a_3(R-R_0)^3|m\rangle\langle m|(R-R_0)|0\rangle}{m\omega}
\end{equation}
Using the theory of ladder operators, $R-R_0=\frac{1}{\sqrt{2\mu\omega}}(\hat{a}^{\dagger}+\hat{a})$, the final result can be obtained:
\begin{equation}
    \langle R\rangle=R_0-\frac{3a_3}{4\mu^2\omega^3}=R_0-\frac{3a_3}{2^{\frac72}a_2^{\frac32}\mu^{\frac12}}
\end{equation}
This result shows that the lowest-order anharmonic correction term is proportional to the inverse square root of the nuclear mass. Additionally, the figure of the PES confirms that $a_3<0$, so the correction term is generally positive.
\clearpage
\section{Calculation of the Contact Spin Density}
In PermNet and Ferminet, we can directly sample wave function configurations to statistically determine the true two-body density $\rho(\mathbf{r})$. Due to the severe inefficiency of Monte Carlo sampling as $\mathbf{r} = |\mathbf{r}_1 - \mathbf{r}_2| \to 0$, the two-body contact density at zero distance cannot be obtained directly. Therefore, it must be extrapolated from the two-body density measured at finite distances. Here, we choose to compute the average two-body density with distance less than $\mathbf{r}$ (i.e., within a sphere, not a shell), which effectively increases the statistical weight of the data with $\mathbf{r} \to 0$. Due to the cusp condition:
\begin{equation}
\left.\frac{d}{d r} \ln \psi(r)\right|_{r=0} = -m Z,
\label{eq:cusp_condition}
\end{equation}
where $Z$ is the charge of the particle, $m$ is the reduced mass. We have:
\begin{equation}
\psi(r) = \psi(0)\left(1 - m Z r + O(r^{2})\right),
\end{equation}
and since $\rho(r) = |\psi(r)|^{2}$,
\begin{equation}
\rho(r) = \rho_{0}\left(1 - 2 m Z r + O(r^{2})\right).
\label{eq:rho_expansion}
\end{equation}
Defining the average two-body density within a sphere of radius $r$ as:
\begin{equation}
\rho_{\text{ball}}(r) = \frac{3}{r^{3}} \int_{0}^{r} \rho(r') r'^{2} dr'.
\label{eq:rho_ball_def}
\end{equation}
Combining with Eq. (\ref{eq:rho_expansion}) we obtain:
\begin{equation}
\rho_{\text{ball}}(r)  = \rho_{0}\left(1 - \frac{3}{2} m Z r + O(r^{2})\right).
\label{eq:rho_ball_expansion}
\end{equation}
Assuming a polynomial extrapolation formula 
$\rho_{\mathrm{ball}}(r) = a_{0} + a_{1} r + a_{2} r^{2} + \cdots$, 
the coefficients satisfy:
\begin{equation}
a_{0} = \rho_{0}, \quad a_{1}  = -\frac{3}{2} m Z a_{0}.
\label{eq:coeff_constraint}
\end{equation}
This imposes a constraint on the first-order term: $a_1 \simeq -1.5 a_0$. 

\begin{figure}[htbp]
\centering
\includegraphics[width=\linewidth]{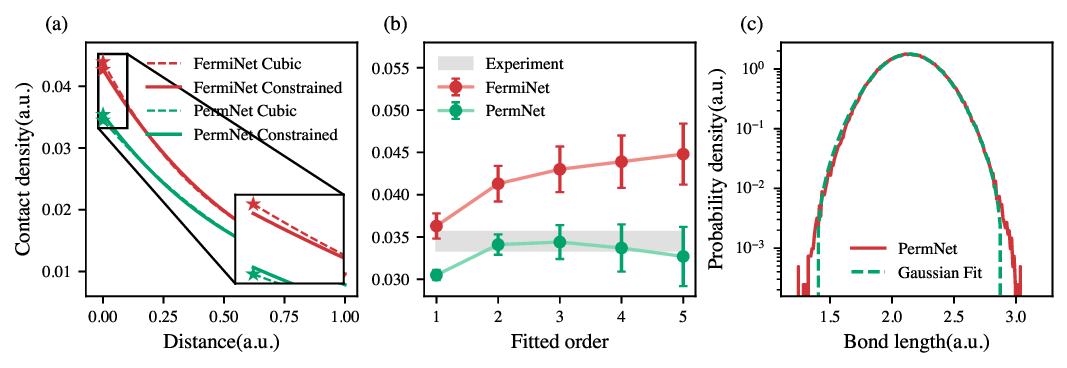}
\caption{(a) The contact spin density fitting curve calculated by FermiNet and PermNet using a cubic polynomial fitting function. The dashed line represents the fitting process without any constraint on the first-order term. The solid line represents the fitting process that satisfies the constraint condition, as in Equation~\eqref{eq:coeff_constraint}. (b) Contact spin density calculated via different fitting order polynomial extrapolation at 500,000 training steps. The fitting process without any constraint on the first-order term. (c) C--Mu distance distribution calculated from PermNet. The green dashed line represent the fitting curve using a Gaussian function.}
\label{fig:s3}
\end{figure}

The extrapolations performed with and without this constraint are shown Fig.~\ref{fig:s3}(a). The hyperfine couplings obtained via the unconstrained extrapolation are shown in Fig.~\ref{fig:s3}(b), while the constrained results are reserved for the main text (Fig. 4).
Given that the constraint has a slight effect on the hyperfine couplings, we therefore present the constrained results as the primary finding in the main text.

\clearpage
\section{Muon Spatial Distribution: A Comparison of PermNet and the Harmonic Approximation}

This section compares the muon spatial distribution obtained from the fully quantized PermNet with that from the BO-harmonic approximation. Since PermNet accounts for molecular translation and overall rotation, we present the relative positions by evaluating the C–Mu distance via its bond length distribution, as shown in Fig.~\ref{fig:s3}(c).

Comparing the C–Mu bond length distributions between PermNet and the BO-harmonic approximation serves to validate our calculations and estimate the strength of anharmonic effects. Under the BO-harmonic approximation, phonon calculations for muoniated ethylene were performed by solving the following equations:
\begin{equation}
\tilde{\mathbf{H}} \mathbf{v}_{k} = \lambda_{k} \mathbf{v}_{k},
\label{eq:eigen_problem}
\end{equation}
where $\tilde{\mathbf{H}}$ is the mass-weighted Hessian matrix:
\begin{equation}
\tilde{H}_{ij} = \frac{1}{\sqrt{m_i m_j}} H_{ij}, \quad H_{ij} = \frac{\partial^2 E_{\mathrm{tot}}}{\partial x_i \partial x_j}.
\label{eq:hessian_def}
\end{equation}
The angular frequency for each vibrational mode is obtained via $\omega_k = \sqrt{\lambda_k}$. The calculations were performed using the Gaussian software with the UB3LYP functional, and the Hessian matrix was obtained analytically. The results are shown in Table~\ref{tab:vibrational_modes}.

\begin{table}[htbp]
\centering
\caption{Vibrational modes of muoniated ethylene obtained under the BO-harmonic approximation. $M_{\nu}$ is the reduced mass of the vibrational mode. $D=\bm{w}_\text{Mu}\cdot\hat{\bm{r}}_\text{C-Mu}$, $\bm{w}_\text{Mu}$ is the vibrational weight of Mu.}
\label{tab:vibrational_modes}
\begin{tabular}{cccc}
\toprule
Mode & $\nu$ (cm$^{-1}$) & $M_{\nu}$ ({a.u.}) & $D$\\
\midrule
1  & 105.9576  & 0.7919 & 0.00\\
2  & 496.1460  & 1.2232 & -0.02\\
3  & 830.1919  & 0.9897 & 0.00\\
4  & 1060.5878 & 2.6761 & -0.02\\
5  & 1142.7607 & 1.2187 & 0.11\\
6  & 1214.7514 & 1.3165 & 0.00\\
7  & 1455.8200 & 1.1047 & 0.01\\
8  & 1477.2494 & 1.1712 & 0.01\\
9  & 3024.5570 & 0.9261 & -0.02\\
10 & 3038.8369 & 0.5620 & 0.00\\
11 & 3140.0073 & 1.0503 & 0.00\\
12 & 3239.1521 & 1.0612 & 0.00\\
13 & 3334.0719 & 0.1398 & 0.00\\
14 & 3391.7190 & 0.1195 & 0.01 \\
15 & 8519.8272 & 0.1143 & -1.00 \\
\bottomrule
\end{tabular}
\end{table}

Mode 15 is a stretching mode where the muon moves along the C--Mu bond direction, allowing for a direct comparison with the PermNet C--Mu bond length distribution (other muon-localized modes lack motion along the C--Mu bond direction, see $D$ in Table \ref{tab:vibrational_modes}). Within the harmonic phonon model, the Gaussian width $\sigma$ at 0~K is given by $\sigma = \sqrt{\hbar/(2 \omega M_{\nu})}$, where $M_{\nu}$ is the reduced mass of the vibrational mode. For mode 15, this yields $\sigma \approx 0.248$~a.u. In comparison, Gaussian fitting of the distribution in Fig.~\ref{fig:s1}(c) gives $\sigma = 0.2261(8)$~a.u. This rough agreement supports the validity of the PermNet calculation. In consideration of the slight errors from the functional, the observed difference can be primarily attributed to the strength of anharmonic effects.

\FloatBarrier